\pdfoutput=1
\documentclass[12pt,a4paper]{article}
\usepackage{ifthen} % for conditional statements
\newboolean{pdflatex}
\setboolean{pdflatex}{true} % False for eps figures 

\newboolean{articletitles}
\setboolean{articletitles}{true} % False removes titles in references

\newboolean{uprightparticles}
\setboolean{uprightparticles}{false} %True for upright particle symbols

\newboolean{inbibliography}
\setboolean{inbibliography}{false} %True once you enter the bibliography

\usepackage{microtype}
\usepackage{lineno}  % for line numbering during review
\usepackage{xspace} % To avoid problems with missing or double spaces after
\usepackage{caption} %these three command get the figure and table captions automatically small

\usepackage{graphicx}  % to include figures (can also use other packages)
\usepackage{color}
\usepackage{colortbl}
\graphicspath{{./figs/}} % Make Latex search fig subdir for figures

\usepackage{amsmath} % Adds a large collection of math symbols
\usepackage{amssymb}
\usepackage{amsfonts}
\usepackage{upgreek} % Adds in support for greek letters in roman typeset

\newcommand*\patchAmsMathEnvironmentForLineno[1]{%
\expandafter\let\csname old#1\expandafter\endcsname\csname #1\endcsname
\expandafter\let\csname oldend#1\expandafter\endcsname\csname
end#1\endcsname
 \renewenvironment{#1}%
   {\linenomath\csname old#1\endcsname}%
   {\csname oldend#1\endcsname\endlinenomath}%
}
\newcommand*\patchBothAmsMathEnvironmentsForLineno[1]{%
  \patchAmsMathEnvironmentForLineno{#1}%
  \patchAmsMathEnvironmentForLineno{#1*}%
}
\AtBeginDocument{%
\patchBothAmsMathEnvironmentsForLineno{equation}%
\patchBothAmsMathEnvironmentsForLineno{align}%
\patchBothAmsMathEnvironmentsForLineno{flalign}%
\patchBothAmsMathEnvironmentsForLineno{alignat}%
\patchBothAmsMathEnvironmentsForLineno{gather}%
\patchBothAmsMathEnvironmentsForLineno{multline}%
\patchBothAmsMathEnvironmentsForLineno{eqnarray}%
}

\usepackage{hyperref}    % Hyperlinks in references
\usepackage[all]{hypcap} % Internal hyperlinks to floats.

\def\lhcb {\mbox{LHCb}\xspace}

\def\MagUp {\mbox{\em Mag\kern -0.05em Up}\xspace}

\ifthenelse{\boolean{uprightparticles}}%
{

 \def\Ppi         {\ensuremath{\uppi}\xspace}

 \def\Ppsi        {\ensuremath{\uppsi}\xspace}

 \def\PDelta      {\ensuremath{\Delta}\xspace}                 
 \def\PXi      {\ensuremath{\Xi}\xspace}                 
 \def\PLambda      {\ensuremath{\Lambda}\xspace}                 
 \def\PSigma      {\ensuremath{\Sigma}\xspace}                 
 \def\POmega      {\ensuremath{\Omega}\xspace}                 
 \def\PUpsilon      {\ensuremath{\Upsilon}\xspace}

 \def\PB      {\ensuremath{\mathrm{B}}\xspace}                 
                  
 \def\PD      {\ensuremath{\mathrm{D}}\xspace}

 \def\PJ      {\ensuremath{\mathrm{J}}\xspace}                 
 \def\PK      {\ensuremath{\mathrm{K}}\xspace}

 \def\Pb      {\ensuremath{\mathrm{b}}\xspace}                 
 \def\Pc      {\ensuremath{\mathrm{c}}\xspace}

 \def\Pi      {\ensuremath{\mathrm{i}}\xspace}

 \def\Ps      {\ensuremath{\mathrm{s}}\xspace}

}
{

 \def\Ppi         {\ensuremath{\pi}\xspace}

 \def\Ppsi        {\ensuremath{\psi}\xspace}                 
                  
 \mathchardef\PDelta="7101
 \mathchardef\PXi="7104
 \mathchardef\PLambda="7103
 \mathchardef\PSigma="7106
 \mathchardef\POmega="710A
 \mathchardef\PUpsilon="7107
                  
 \def\PB      {\ensuremath{B}\xspace}                 
                  
 \def\PD      {\ensuremath{D}\xspace}

 \def\PJ      {\ensuremath{J}\xspace}                 
 \def\PK      {\ensuremath{K}\xspace}

 \def\Pb      {\ensuremath{b}\xspace}                 
 \def\Pc      {\ensuremath{c}\xspace}

 \def\Pi      {\ensuremath{i}\xspace}

 \def\Ps      {\ensuremath{s}\xspace}

}

\makeatletter
\ifcase \@ptsize \relax% 10pt
  \newcommand{\miniscule}{\@setfontsize\miniscule{4}{5}}% \tiny: 5/6
\or% 11pt
  \newcommand{\miniscule}{\@setfontsize\miniscule{5}{6}}% \tiny: 6/7
\or% 12pt
  \newcommand{\miniscule}{\@setfontsize\miniscule{5}{6}}% \tiny: 6/7
\fi
\makeatother

\DeclareRobustCommand{\optbar}[1]{\shortstack{{\miniscule (\rule[.5ex]{1.25em}{.18mm})}
  \\ [-.7ex] $#1$}}

\def\squark    {{\ensuremath{\Ps}}\xspace}

\def\cquark    {{\ensuremath{\Pc}}\xspace}

\def\bquark    {{\ensuremath{\Pb}}\xspace}

\def\pion   {{\ensuremath{\Ppi}}\xspace}
\def\piz    {{\ensuremath{\pion^0}}\xspace}

\def\pip    {{\ensuremath{\pion^+}}\xspace}
\def\pim    {{\ensuremath{\pion^-}}\xspace}

\def\kaon    {{\ensuremath{\PK}}\xspace}
  \def\Kbar    {{\kern 0.2em\overline{\kern -0.2em \PK}{}}\xspace}

\def\KorKbar    {\kern 0.18em\optbar{\kern -0.18em K}{}\xspace}

\def\Kp      {{\ensuremath{\kaon^+}}\xspace}
\def\Km      {{\ensuremath{\kaon^-}}\xspace}

\def\KS      {{\ensuremath{\kaon^0_{\rm\scriptscriptstyle S}}}\xspace}

  \def\Dbar    {{\kern 0.2em\overline{\kern -0.2em \PD}{}}\xspace}
\def\D       {{\ensuremath{\PD}}\xspace}

\def\DorDbar    {\kern 0.18em\optbar{\kern -0.18em D}{}\xspace}
\def\Dz      {{\ensuremath{\D^0}}\xspace}
\def\Dzb     {{\ensuremath{\Dbar{}^0}}\xspace}

\def\DorDstarz  {\ensuremath{\D{}^{(*)0}}\xspace}
\def\Dstarzb {{\ensuremath{\Dbar{}^{*0}}}\xspace}
\def\DorDstarzb {\ensuremath{\Dbar{}^{(*)0}}\xspace}

\def\Dstarm  {{\ensuremath{\D^{*-}}}\xspace}

\def\B       {{\ensuremath{\PB}}\xspace}
\def\Bbar    {{\ensuremath{\kern 0.18em\overline{\kern -0.18em \PB}{}}}\xspace}

\def\BorBbar    {\kern 0.18em\optbar{\kern -0.18em B}{}\xspace}
\def\Bz      {{\ensuremath{\B^0}}\xspace}
\def\Bzb     {{\ensuremath{\Bbar{}^0}}\xspace}
\def\Bu      {{\ensuremath{\B^+}}\xspace}

\def\Bd      {{\ensuremath{\B^0}}\xspace}
\def\Bs      {{\ensuremath{\B^0_\squark}}\xspace}
\def\Bsb     {{\ensuremath{\Bbar{}^0_\squark}}\xspace}

\def\Bds     {\ensuremath{\B^0_{(\squark)}}\xspace}

\def\jpsi     {{\ensuremath{{\PJ\mskip -3mu/\mskip -2mu\Ppsi\mskip 2mu}}}\xspace}

  \def\Y#1S{\ensuremath{\PUpsilon{(#1S)}}\xspace}% no space before {...}!

\def\Lz          {{\ensuremath{\PLambda}}\xspace}
\def\Lbar        {{\ensuremath{\kern 0.1em\overline{\kern -0.1em\PLambda}}}\xspace}
\def\LorLbar    {\kern 0.18em\optbar{\kern -0.18em \PLambda}{}\xspace}

\def\Lb      {{\ensuremath{\Lz^0_\bquark}}\xspace}

\def\Lc      {{\ensuremath{\Lz^+_\cquark}}\xspace}

\def\to                 {\ensuremath{\rightarrow}\xspace}

\def\CP                {{\ensuremath{C\!P}}\xspace}

\def\AT#1     {\ensuremath{A_{\mathrm{T}}^{#1}}\xspace}           % 2

\def\C#1      {\ensuremath{\mathcal{C}_{#1}}\xspace}                       % 9
\def\Cp#1     {\ensuremath{\mathcal{C}_{#1}^{'}}\xspace}                    % 7
\def\Ceff#1   {\ensuremath{\mathcal{C}_{#1}^{\mathrm{(eff)}}}\xspace}        % 9  
\def\Cpeff#1  {\ensuremath{\mathcal{C}_{#1}^{'\mathrm{(eff)}}}\xspace}       % 7
\def\Ope#1    {\ensuremath{\mathcal{O}_{#1}}\xspace}                       % 2
\def\Opep#1   {\ensuremath{\mathcal{O}_{#1}^{'}}\xspace}                    % 7

\newcommand{\tev}{\ifthenelse{\boolean{inbibliography}}{\ensuremath{~T\kern -0.05em eV}\xspace}{\ensuremath{\mathrm{\,Te\kern -0.1em V}}}\xspace}
\newcommand{\gev}{\ensuremath{\mathrm{\,Ge\kern -0.1em V}}\xspace}
\newcommand{\mev}{\ensuremath{\mathrm{\,Me\kern -0.1em V}}\xspace}
\newcommand{\kev}{\ensuremath{\mathrm{\,ke\kern -0.1em V}}\xspace}
\newcommand{\ev}{\ensuremath{\mathrm{\,e\kern -0.1em V}}\xspace}
\newcommand{\gevc}{\ensuremath{{\mathrm{\,Ge\kern -0.1em V\!/}c}}\xspace}
\newcommand{\mevc}{\ensuremath{{\mathrm{\,Me\kern -0.1em V\!/}c}}\xspace}
\newcommand{\gevcc}{\ensuremath{{\mathrm{\,Ge\kern -0.1em V\!/}c^2}}\xspace}
\newcommand{\gevgevcccc}{\ensuremath{{\mathrm{\,Ge\kern -0.1em V^2\!/}c^4}}\xspace}
\newcommand{\mevcc}{\ensuremath{{\mathrm{\,Me\kern -0.1em V\!/}c^2}}\xspace}

\def\mum  {\ensuremath{{\,\upmu\rm m}}\xspace}

\def\invfb   {\ensuremath{\mbox{\,fb}^{-1}}\xspace}

\def\gsim{{~\raise.15em\hbox{$>$}\kern-.85em
          \lower.35em\hbox{$\sim$}~}\xspace}
\def\lsim{{~\raise.15em\hbox{$<$}\kern-.85em
          \lower.35em\hbox{$\sim$}~}\xspace}

\def\ptot       {\mbox{$p$}\xspace}
\def\pt         {\mbox{$p_{\rm T}$}\xspace}

\def\evtgen     {\mbox{\textsc{EvtGen}}\xspace}

\def\geant      {\mbox{\textsc{Geant4}}\xspace}

\def\photos     {\mbox{\textsc{Photos}}\xspace}

\def\pythia     {\mbox{\textsc{Pythia}}\xspace}

\def\tell1  {TELL1\xspace}
\def\ukl1   {UKL1\xspace}

\usepackage{cite} % Allows for ranges in citations
\usepackage{mciteplus}
\usepackage{longtable} % only for template; not usually to be used in PAPERs

\begin{document}
\begin{titlepage}
\pagenumbering{roman}

% Header ---------------------------------------------------
\vspace*{-1.5cm}
\centerline{\large EUROPEAN ORGANIZATION FOR NUCLEAR RESEARCH (CERN)}
\vspace*{1.5cm}
\hspace*{-0.5cm}
\begin{tabular*}{\linewidth}{lc@{\extracolsep{\fill}}r}
\ifthenelse{\boolean{pdflatex}}% Logo format choice
{\vspace*{-2.7cm}\mbox{\!\!\!\includegraphics[width=.14\textwidth]{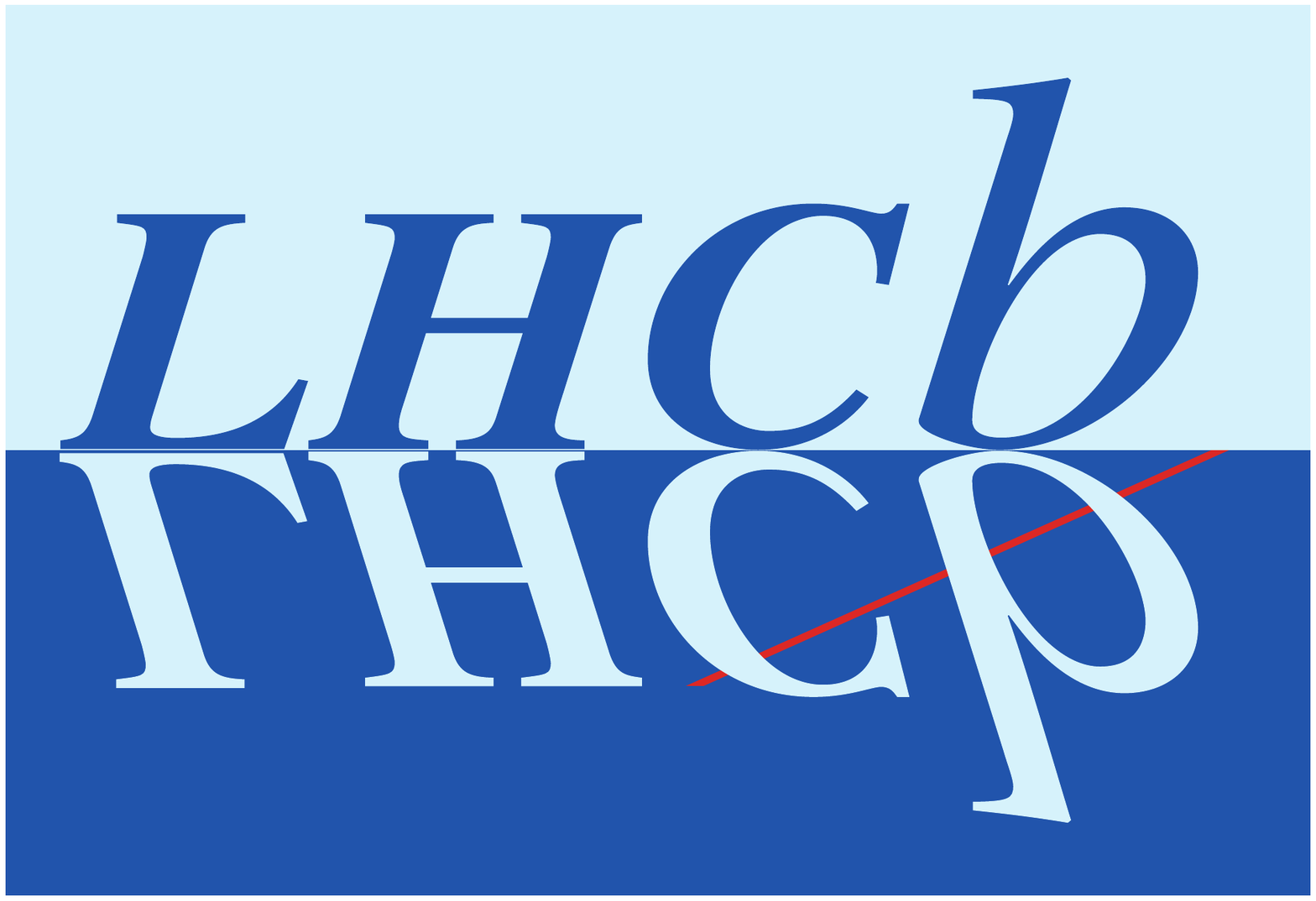}} & &}%
{\vspace*{-1.2cm}\mbox{\!\!\!\includegraphics[width=.12\textwidth]{lhcb-logo.eps}} & &}%
\\
 & & CERN-PH-EP-2015-086 \\  % ID 
 & & LHCb-PAPER-2015-012 \\  % ID 
 & & May 7, 2015 \\ % Date - Can also hardwire e.g.: 23 March 2010
 & & \\
% not in paper \hline
\end{tabular*}

\vspace*{2.5cm}

% Title --------------------------------------------------
{\bf\boldmath\huge
\begin{center}
  Search for the decay $B_s^0 \to \Dzb f_{0}(980)$
\end{center}
}

\vspace*{1.5cm}

% Authors -------------------------------------------------
\begin{center}
The LHCb collaboration\footnote{Authors are listed at the end of this paper.}
\end{center}

\vspace{\fill}

% Abstract -----------------------------------------------
\begin{abstract}
  \noindent
  A search for $B_s^0 \to \Dzb f_{0}(980)$ decays is performed using $3.0\invfb$ of $pp$ collision data recorded by the LHCb experiment during 2011 and 2012.
  The $f_{0}(980)$ meson is reconstructed through its decay to the $\pip\pim$ final state in the mass window $900 \mevcc < m(\pip\pim) < 1080 \mevcc$.
  No significant signal is observed.
  The first upper limits on the branching fraction of 
  $\mathcal{B}(B_s^0 \to \Dzb f_{0}(980)) < 3.1\,(3.4) \times 10^{-6}$ 
  are set at $90\,\%$ ($95\,\%$) confidence level.
\end{abstract}

\vspace*{1.5cm}

\begin{center}
  Published in JHEP
\end{center}

\vspace{\fill}

{\footnotesize 
\centerline{\copyright~CERN on behalf of the \lhcb collaboration, licence \href{http://creativecommons.org/licenses/by/4.0/}{CC-BY-4.0}.}}
\vspace*{2mm}

\end{titlepage}

%  empty page follows the title page ----
\newpage
\setcounter{page}{2}
\mbox{~}

\cleardoublepage
% %%%%%%%%%%%%% ---------

\renewcommand{\thefootnote}{\arabic{footnote}}
\setcounter{footnote}{0}

%%%%%%%%%%%%%%%%%%%%%%%%%
%%%%% Main text %%%%%%%%%
%%%%%%%%%%%%%%%%%%%%%%%%%

\pagestyle{plain} % restore page numbers for the main text
\setcounter{page}{1}
\pagenumbering{arabic}

%\linenumbers

\section{Introduction}
\label{sec:introduction}

Understanding the quark-level substructure of the scalar mesons is one of the main challenges in hadronic physics.  
The number of observed states, and their masses and branching fractions, suggest that there is a contribution from four-quark wavefunctions in addition to $q\bar{q}$, and possible gluonic, degrees of freedom~\cite{mesons,PDG2014}.
However, the extent of mixing between the different components is unclear.

Measurement of the relative production of scalar mesons in $\Bd$ and $\Bs$ meson decays can help to address this issue~\cite{Fleischer:2011au,Stone:2013eaa}.
Measurements of $\Bds\to\jpsi f$ decays, where $f$ represents either the $f_0(500)$ (also known as $\sigma$) or the $f_0(980)$ meson, and $f \to \pip\pim$~\cite{LHCb-PAPER-2012-005,LHCb-PAPER-2012-045,LHCb-PAPER-2013-069,LHCb-PAPER-2014-012} have already provided important insight into the structure of the scalar mesons~\cite{Liang:2014tia,Close:2015rza}.
Studies of $\Bds\to \Dzb f$ decays provide complementary information to the $\Bds\to\jpsi f$ case~\cite{Liang:2014ama}.
Measurements of the branching fractions of $\Bd \to \Dzb f_0(500)$ and $\Bd \to \Dzb f_0(980)$ decays have been obtained from Dalitz plot analyses of $\Bd \to \Dzb \pip\pim$ decays~\cite{Kuzmin:2006mw,LHCb-PAPER-2014-070}, but there is no experimental result to date on the $\Bs$ decays.

In addition, under the assumption that the $f_0(980)$ meson has a predominant $s\bar{s}$ component, the $\Bs\to \Dzb f_0(980)$ decay mode can be used to determine the angle $\gamma$ of the CKM unitarity triangle~\cite{Cabibbo:1963yz,Kobayashi:1973fv}, using the same methods that are applicable for the $\Bs\to\Dzb\phi$ decay mode~\cite{Gronau:1990ra,Gronau:2004gt,Gronau:2007bh,Nandi:2011uw,Wang:2011zw}.
Since the $\Bs\to\Dzb\phi$ decay has recently been observed~\cite{LHCb-PAPER-2013-035}, a signal for the $\Bs\to \Dzb f_0(980)$ channel is expected if the branching fractions of the two decays are comparable.
An explicit calculation predicts ${\cal B}\left(\Bs\to \Dzb f_0(980)\right) = \left(3.50\,^{+1.26}_{-1.15}\,^{+0.56}_{-0.77}\right)\times 10^{-5}$~\cite{Kim:2013ria}.

In this paper, the result of a search for the $\Bs\to\Dzb f_0(980)$ decay is presented. 
The inclusion of charge conjugated processes is implied throughout the paper. 
The final state is reconstructed through the $\Dzb \to \Kp\pim$ and $f_0(980) \to \pip\pim$ decays.
The decay-time-integrated branching fraction is measured under the assumption that the $\Bs \to \Dzb \pip\pim$ decay proceeds uniquely via the $f_0(980)$ resonance within the selected mass window, $900 \mevcc < m(\pip\pim) < 1080 \mevcc$.
This approach was used for the first observation of $\Bs\to\jpsi f_0(980)$ decays~\cite{LHCb-PAPER-2011-002}; it is also justified by the fact that no other contribution to $\Bs \to \Dzb \pip\pim$ decays, for example through the $\Bs \to \Dstarm \pip$ process~\cite{LHCb-PAPER-2012-056}, is expected at the current level of sensitivity.
A further assumption is that the contribution from $\Bs\to\Dz f_0(980)$ decays, which is suppressed by the ratio of CKM matrix elements $\left| V^{}_{ub}V^{*}_{cs}/ (V^{}_{cb}V^{*}_{us}) \right|^2 \approx 0.1$, is negligible. 
Formally, the measurement is of the decay-time-integrated sum of the branching fractions for $\Bs\to\Dzb f_0(980)$ and $\Bs\to\Dz f_0(980)$ decays.

The analysis is based on $3.0\invfb$ of LHC $pp$ collision data collected with the LHCb detector, with approximately one third taken at a centre-of-mass energy of 7\tev during 2011 and the remainder at 8\tev during 2012. 
The measurement is obtained by evaluating the ratio of branching fractions
\begin{equation}\label{eqn:bf}
\frac{\mathcal{B}(\Bs\to\Dzb f_0(980))}{\mathcal{B}(\Bd\to\Dzb\pip\pim)} = \frac{N(\Bs\to\Dzb f_0(980))}{N(\Bd\to\Dzb\pip\pim)} \times
\frac{\epsilon(\Bd\to\Dzb\pip\pim)}{\epsilon(\Bs\to\Dzb f_0(980))} \times \frac{f_d}{f_s}\,,
\end{equation}
from which the absolute branching fraction for $\Bs\to\Dzb f_0(980)$ decays is determined using the known value of $\mathcal{B}(\Bd\to\Dzb\pip\pim)$~\cite{LHCb-PAPER-2014-070}.
The yields $N(\Bs\to\Dzb f_0(980))$ and $N(\Bd\to\Dzb\pip\pim)$ are determined from separate extended maximum likelihood fits to the distributions of selected $\Dzb f_0(980)$ and $\Dzb\pip\pim$ candidates in both the \B candidate mass and the output of a neural network (NN) used to separate signal from combinatorial background.
The combined reconstruction and selection efficiencies, $\epsilon(\Bs\to\Dzb f_0(980))$ and $\epsilon(\Bd\to\Dzb\pip\pim)$, are determined from simulated samples with data-driven corrections applied.
The ratio of fragmentation fractions inside the LHCb acceptance has been measured to be $f_s/f_d = 0.259 \pm 0.015$~\cite{fsfd}.
Equation~(\ref{eqn:bf}) corresponds to a branching fraction for $f_0(980) \to \pip\pim$ of $100\,\%$, which is the conventional way to quote results for decays involving $f_0(980)$ mesons.

\section{LHCb detector}
\label{sec:detector}

The \lhcb detector~\cite{Alves:2008zz,LHCb-DP-2014-002} is a single-arm forward
spectrometer covering the \mbox{pseudorapidity} range $2<\eta <5$,
designed for the study of particles containing \bquark or \cquark
quarks. The detector includes a high-precision tracking system
consisting of a silicon-strip vertex detector~\cite{LHCb-DP-2014-001}
surrounding the $pp$ interaction region, a large-area silicon-strip detector
located upstream of a dipole magnet with a bending power of about
$4{\rm\,Tm}$, and three stations of silicon-strip detectors and straw
drift tubes~\cite{LHCb-DP-2013-003} placed downstream of the magnet.
The tracking system provides a measurement of momentum, \ptot, of charged particles with
a relative uncertainty that varies from 0.5\% at low momentum to 1.0\% at 200\gevc.
The minimum distance of a track to a primary vertex, the impact parameter (IP), is measured with a resolution of $(15+29/\pt)\mum$,
where \pt is the component of the momentum transverse to the beam, in \gevc.
Different types of charged hadrons are distinguished using information
from two ring-imaging Cherenkov detectors~\cite{LHCb-DP-2012-003}. 
Photon, electron and
hadron candidates are identified by a calorimeter system consisting of
scintillating-pad and preshower detectors, an electromagnetic
calorimeter and a hadronic calorimeter. Muons are identified by a
system composed of alternating layers of iron and multiwire
proportional chambers~\cite{LHCb-DP-2012-002}.

The trigger~\cite{LHCb-DP-2012-004} consists of a
hardware stage, based on information from the calorimeter and muon
systems, followed by a software stage, in which all tracks
with $\pt>500~(300)\mev$ are reconstructed for 2011 (2012) data.
The software trigger requires a two-, three- or four-track
secondary vertex with significant displacement from the primary
$pp$ interaction vertices~(PVs). At least one charged particle
must have $\pt > 1.7\gevc$ and be inconsistent with originating from a PV.
A multivariate algorithm~\cite{BBDT} is used for
the identification of secondary vertices consistent with the decay
of a \bquark hadron.

Simulated events are used to characterise the detector response to signal and
certain types of background events.
In the simulation, $pp$ collisions are generated using
\pythia~\cite{Sjostrand:2006za,*Sjostrand:2007gs} with a specific \lhcb
configuration~\cite{LHCb-PROC-2010-056}.  Decays of hadronic particles
are described by \evtgen~\cite{Lange:2001uf}, in which final state
radiation is generated using \photos~\cite{Golonka:2005pn}. The
interaction of the generated particles with the detector and its
response are implemented using the \geant
toolkit~\cite{Allison:2006ve, *Agostinelli:2002hh} as described in
Ref.~\cite{LHCb-PROC-2011-006}.

\section{Selection}
\label{sec:selection}

Candidates consistent with the decay chain $\Bds\to\Dzb\pip\pim$ with $\Dzb\to\Kp\pim$ are selected.
The selection procedure involves applying a preselection to the data sample before using a NN to reduce the combinatorial background. 
The NN~\cite{Feindt:2006pm} is trained with the preselected $\Dzb\pip\pim$ data sample, using the {\it sPlot} method~\cite{Pivk:2004ty} with the \B candidate mass as discriminating variable to separate statistically the signal and background categories. 
The input variables to the NN are related to the kinematic properties of the candidate, its isolation from the rest of the $pp$ collision event, and the topology of the signal decay chain. 
Full details of the preselection and NN training can be found in Ref.~\cite{LHCb-PAPER-2014-036}.
The four final-state tracks must also satisfy particle identification (PID) requirements.
Signal candidates are retained for further analysis if they have invariant mass in the range 5100--5900\mevcc.
A requirement that the NN output is greater than $-0.7$ removes $77\,\%$ of combinatorial background and retains $95\,\%$ of $\Bs\to\Dzb f_0(980)$ decays.  

A requirement $m(\Dzb\pim) > 2.10\gevcc$ is used to remove candidates that predominantly originate from $\Bd\to\Dstarm\pip$ decays.
A further requirement, $m(\Dzb\pip) < 5.14\gevcc$, is used to remove backgrounds from $\Bu\to\Dzb\pip$ decays combined with a random $\pim$ candidate.
This source of combinatorial background is kinematically excluded from the signal region, but causes structure in the mass distribution at higher \B candidate mass.  
A similar contribution from $\Bu\to\Dstarzb\pip$ decays cannot be vetoed in the same way, and must therefore be considered further as a source of background.

Following all selection requirements, approximately $1\,\%$ of events contain more than one candidate.
All candidates are retained for the subsequent analysis; the associated systematic uncertainty is negligible.

\section{Determination of signal yield}
\label{sec:fitting}

The yields of $\Bs\to\Dzb f_0(980)$ and $\Bd\to\Dzb\pip\pim$ decays are obtained from two separate extended maximum likelihood fits to the distributions of NN output and \B candidate mass for selected candidates.
The only difference between the samples used in the two fits is that the former has an additional requirement of $900\mevcc < m(\pip\pim) < 1080\mevcc$.
The yield of $\Bs\to\Dzb\pip\pim$ decays in the latter fit is expected to be negligible compared to the large yields of $\Bd$ decays and combinatorial background, and is therefore fixed to zero.
However, a significant number of $\Bd\to\Dzb\pip\pim$ decays are expected to remain within the $f_0(980)$ mass window~\cite{LHCb-PAPER-2014-070}, and therefore both $\Bd$ and $\Bs$ components are included in the former fit.

The data are divided into five bins of the NN output variable, defined as [$-0.70$, 0.03], [0.03, 0.54], [0.54, 0.77], [0.77, 0.88] and [0.88, 1.00], and referred to hereafter as bins 1 to 5, respectively.
The five bins contain a similar proportion of signal decays and increase in purity from bin 1 to bin 5.
This choice of binning has been found to enhance the sensitivity whilst giving stable fit performance. 

The fits include components due to signal and combinatorial background as well as from partially reconstructed and misidentified $b$-hadron decays. 
The signal invariant mass distribution is described by the sum of two Crystal Ball (CB)~\cite{Skwarnicki:1986xj} functions, with a shared mean and tails on opposite sides described by parameters that are fixed to values found in fits to simulated samples. 

The combinatorial background is modelled with the sum of two components.  
The first has an exponential shape, described by a parameter that is the same in all NN bins.  
The second originates from $\Bu\to\Dstarzb\pip$ decays combined with a random pion candidate, and is modelled using a non-parametric shape determined from simulation.  
The limited sizes of the simulated samples used to obtain this and similar shapes are sources of systematic uncertainty.

Partially reconstructed backgrounds occur from $\Bd\to\Dstarzb\pip\pim$ decays, with $\Dstarzb\to\Dzb\piz$ and $\Dstarzb\to\Dzb\gamma$ where the neutral pion or photon is not associated with the candidate, and from $\Bu\to\Dzb\pip\pim\pip$ decays where one $\pip$ is also not associated with the candidate. 
The invariant mass shapes of these backgrounds are described with non-parametric functions derived from simulation.
A global offset of the shape of the partially reconstructed background is determined from the fit to data to allow for differences between data and simulation~\cite{LHCb-PAPER-2015-007}.

Backgrounds from misidentified $b$-hadron decays arise from $\Bd\to\DorDstarzb \Kp\pim$ and $\Bs\to\DorDstarzb \Km\pip$ (hereafter collectively referred to as $\Bds\to\DorDstarzb K\pi$) decays where the kaon is misidentified as a pion and from $\Lb\to\DorDstarz p\pim$ decays where the proton is misidentified as a pion. 
Simulation is used to obtain non-parametric descriptions of the invariant mass shapes.
To obtain these shapes, the latest knowledge of the phase-space distributions of the decays~\cite{LHCb-PAPER-2013-022,LHCb-PAPER-2013-056,LHCb-PAPER-2014-036}, of the relative branching fractions of the $\Bd$ and $\Bs\to\Dzb K\pi$ modes~\cite{LHCb-PAPER-2013-022}, and of the relative branching fractions of the decays involving $\Dzb$ and $\Dstarzb$ mesons~\cite{PDG2014}, is used. 
Data-driven estimates of the misidentification probability as a function of particle kinematic properties are also included.
The relative yields in the NN bins are taken to be the same as for the signal decays.

\begin{figure}[!tb]
\centering
\includegraphics[scale=0.39]{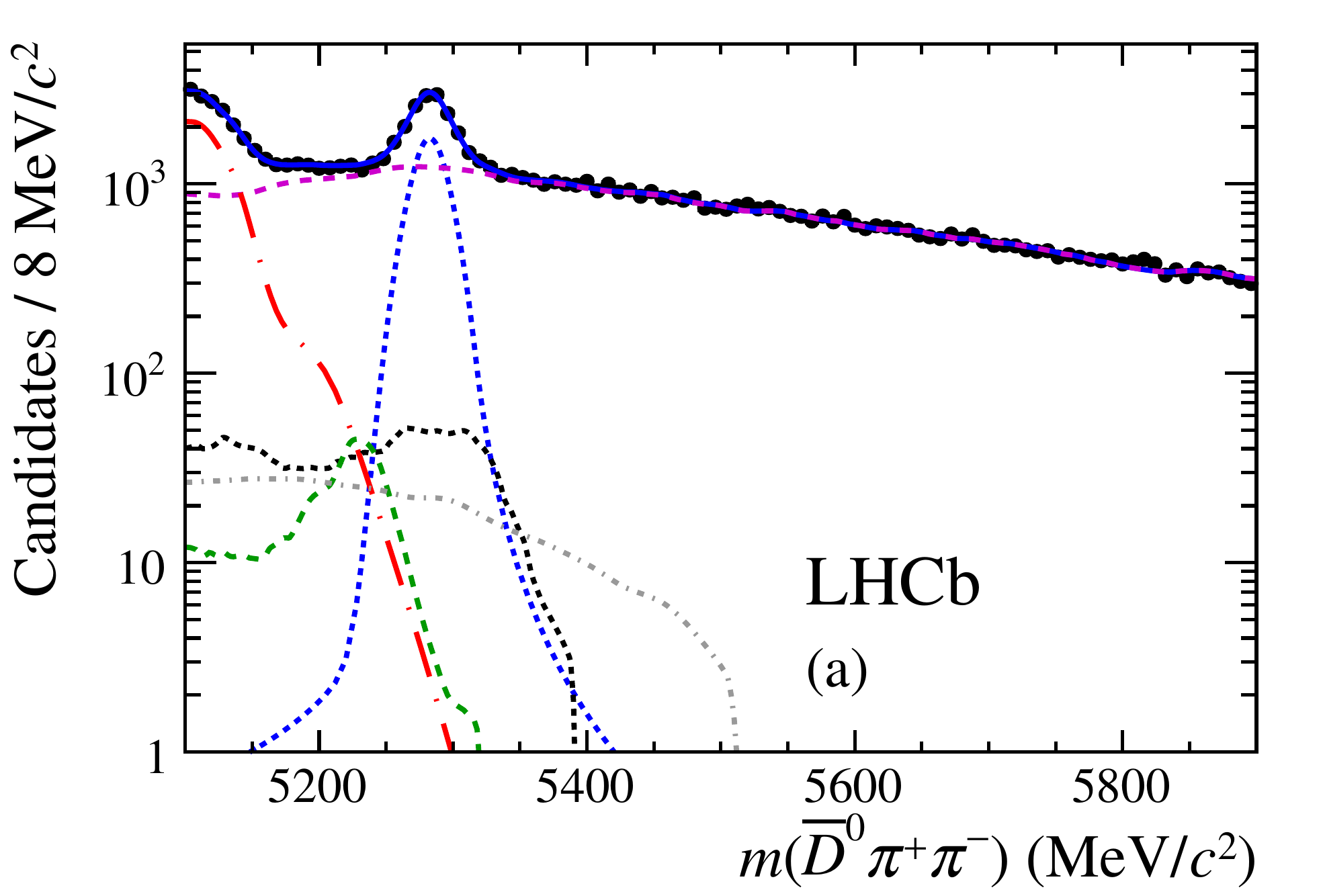}
\includegraphics[scale=0.39]{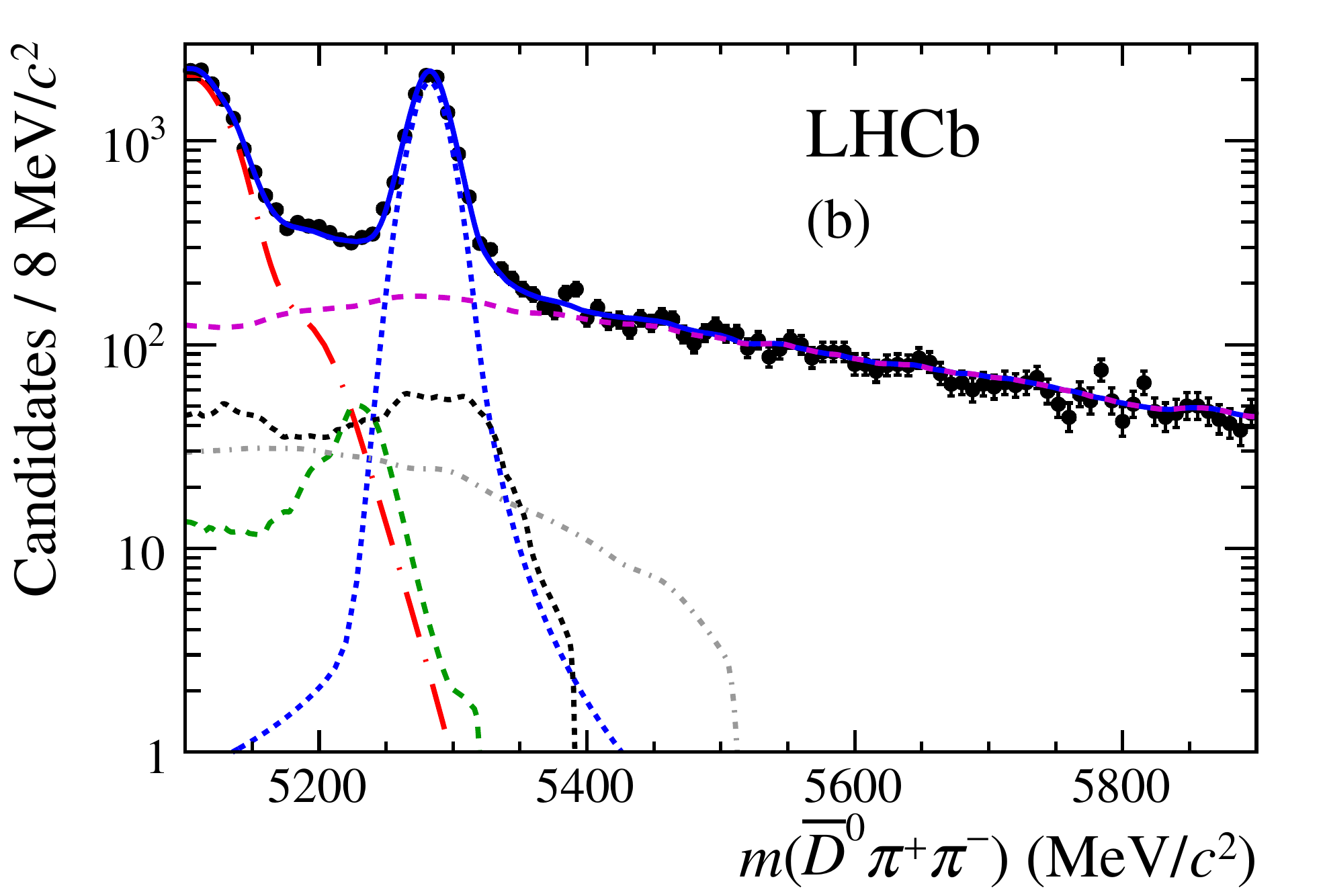}
\includegraphics[scale=0.39]{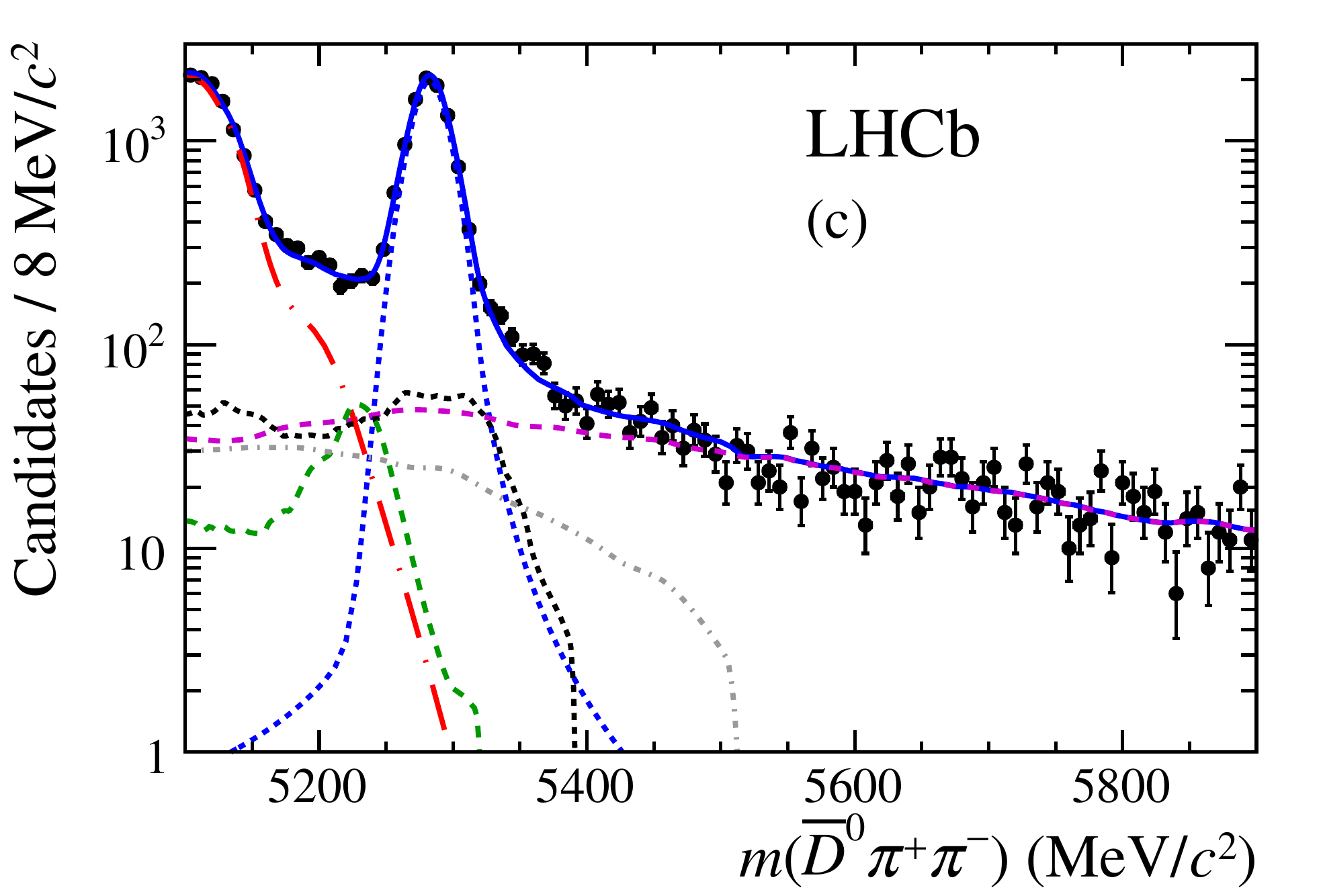}
\includegraphics[scale=0.39]{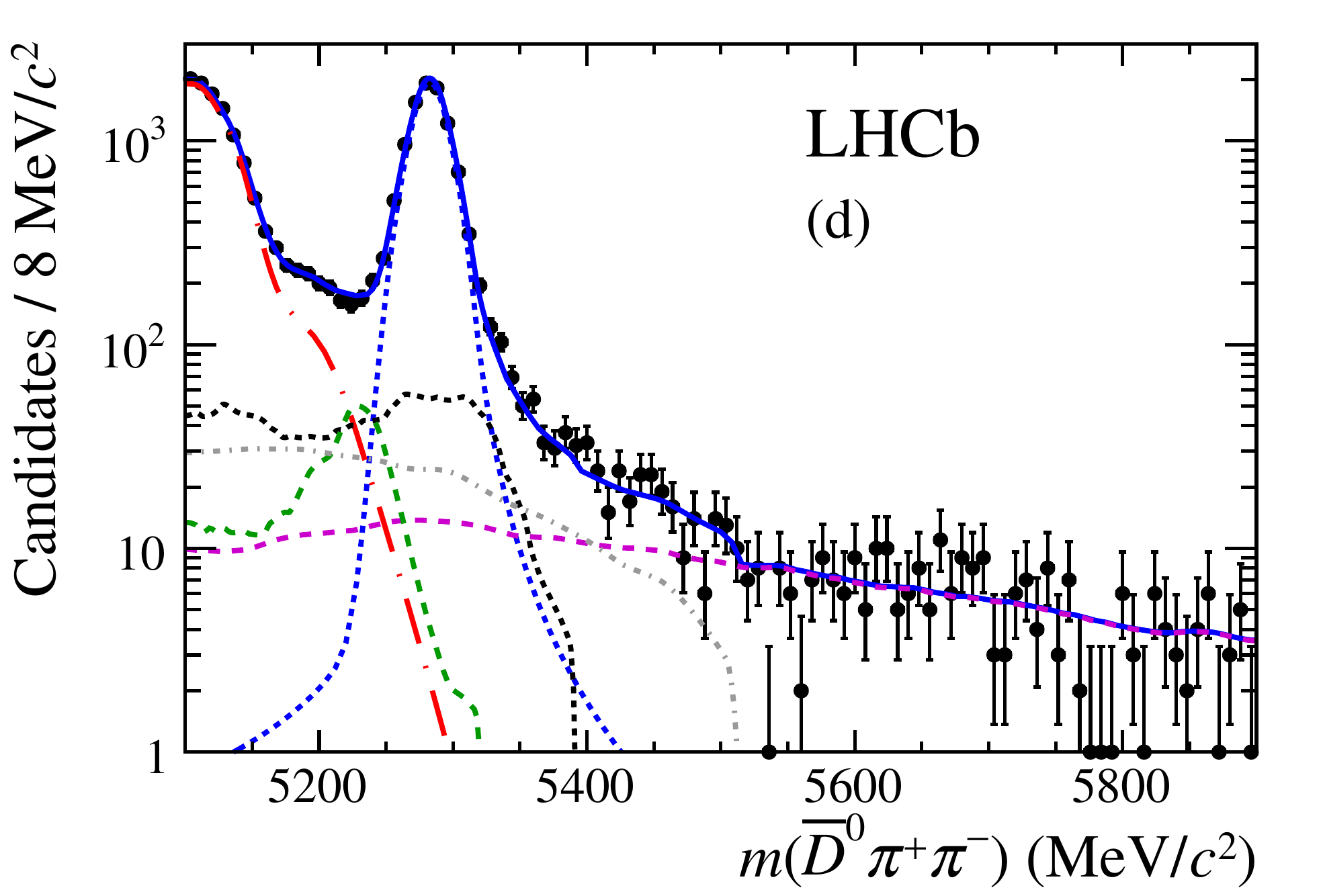}
\includegraphics[scale=0.39]{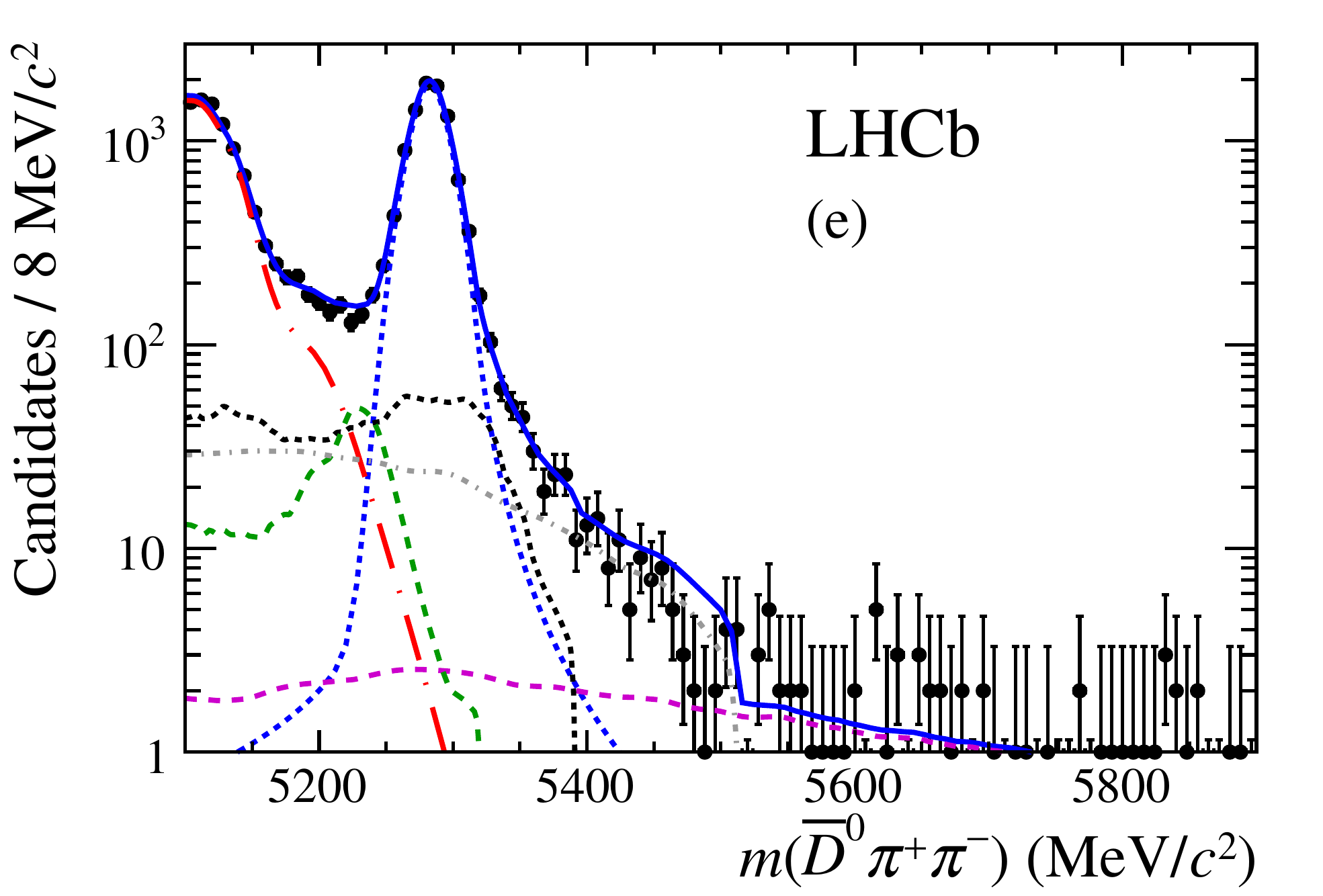}
\includegraphics[scale=0.39]{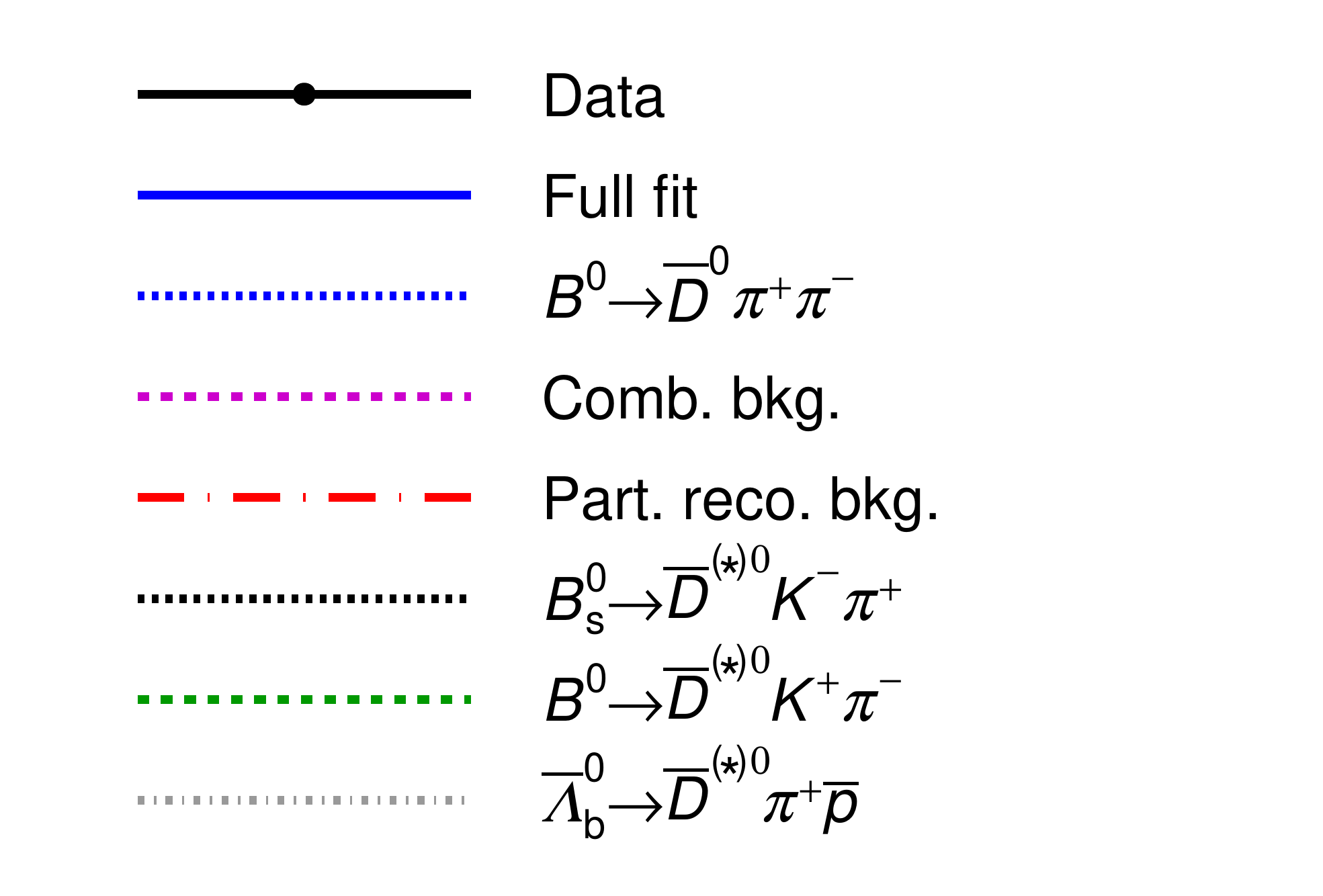}
\caption{\small
  Invariant mass distribution of candidates in the $\Dzb\pip\pim$ data sample with fit results overlaid, shown on a logarithmic scale.
  The components are as detailed in the legend. The labels (a) to (e) show the NN bins with increasing purity.
  The NN binning scheme is described in Sec.~\ref{sec:fitting}.
}
\label{fig:confit}
\end{figure}

A total of 25 parameters are determined from the fit to the $\Dzb\pip\pim$ sample.
These include yields of $\Bd\to\Dzb\pip\pim$ decays, the total combinatorial background, the total partially reconstructed background, and the $\Bds\to\DorDstarzb K\pi$ and $\Lb\to\DorDstarz p\pim$ misidentified backgrounds.
For $\Bd\to\Dzb\pip\pim$ decays, combinatorial and partially reconstructed backgrounds, the fractional yields $f_i$ of each component in bins 1--4 are also free parameters, with the fraction in bin 5 determined as $f_5 = 1-\sum_{i=1}^4 f_{i}$.
In addition, the exponential slope parameter of the combinatorial background, the fraction of the combinatorial background from $\Bu\to\Dstarzb\pip$ decays, the fraction of the partially reconstructed background from $\Bd\to\Dstarzb\pip\pim$ decays and the offset parameter of the partially reconstructed background are determined by the fit.
Parameters of the signal invariant mass shape (the peak position, the width of the core CB function, and the relative normalisation and ratio of the CB widths) are also allowed to vary.
Results of this fit are shown in Fig.~\ref{fig:confit}.

\begin{figure}[!tb]
\centering
\includegraphics[scale=0.39]{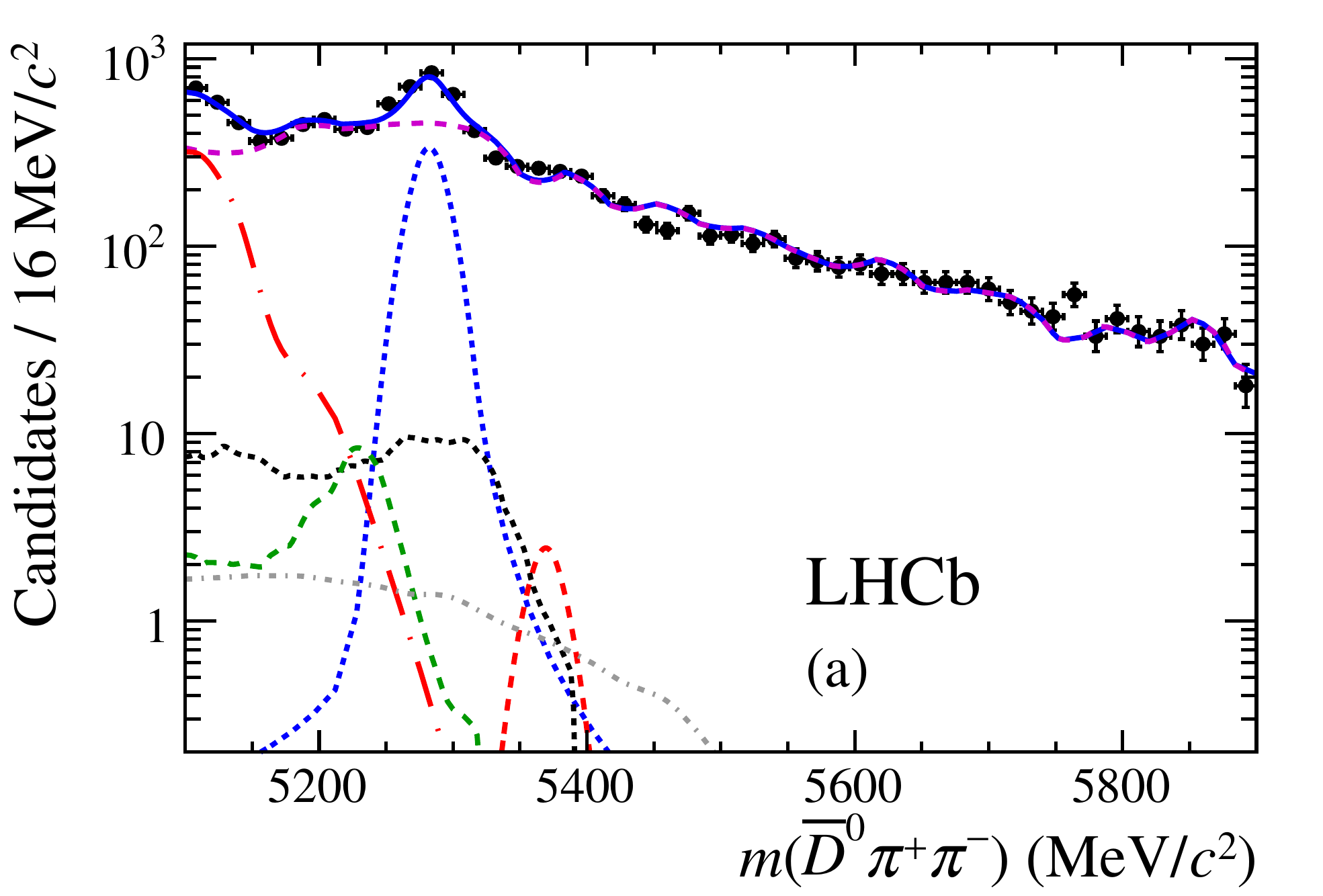}
\includegraphics[scale=0.39]{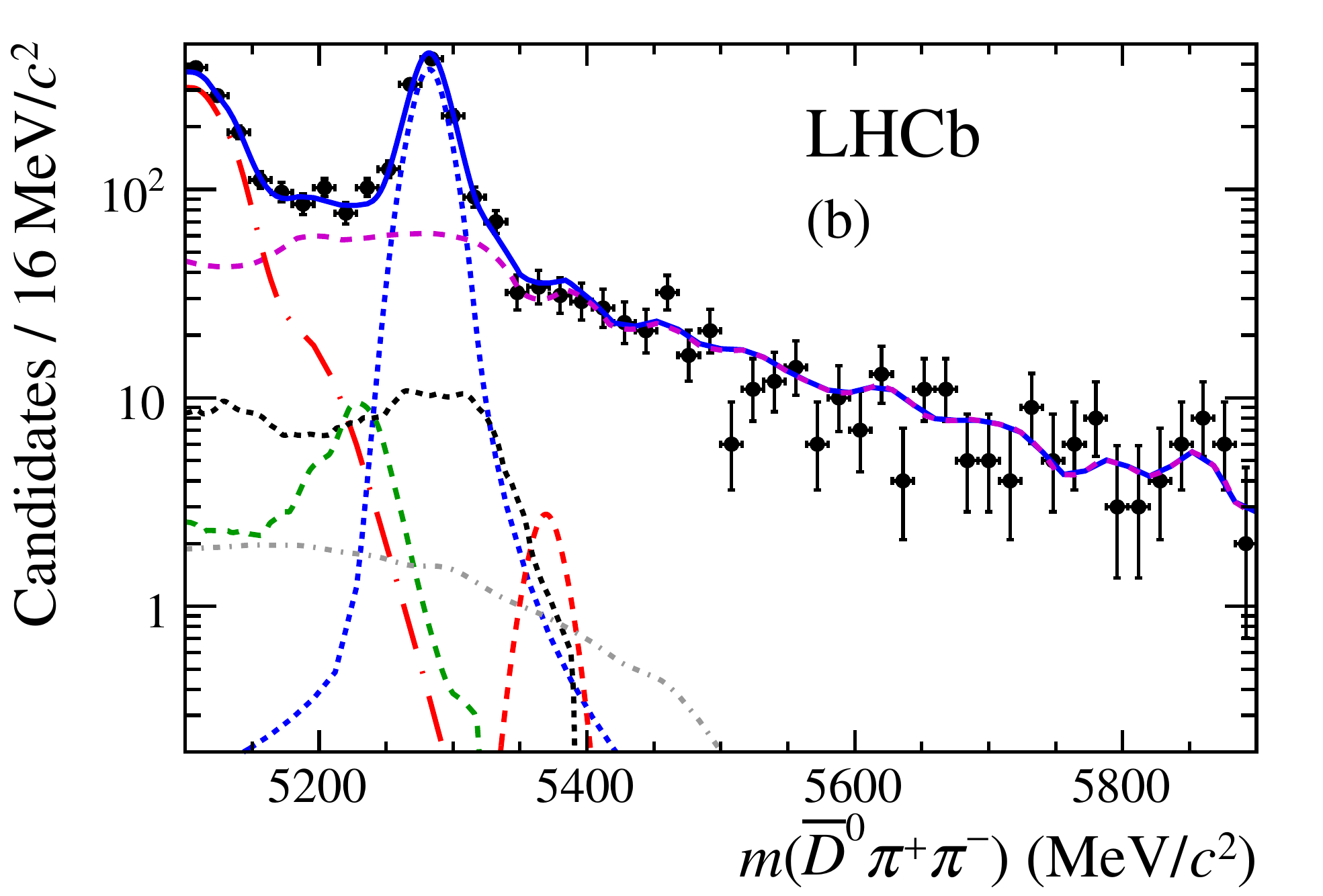}
\includegraphics[scale=0.39]{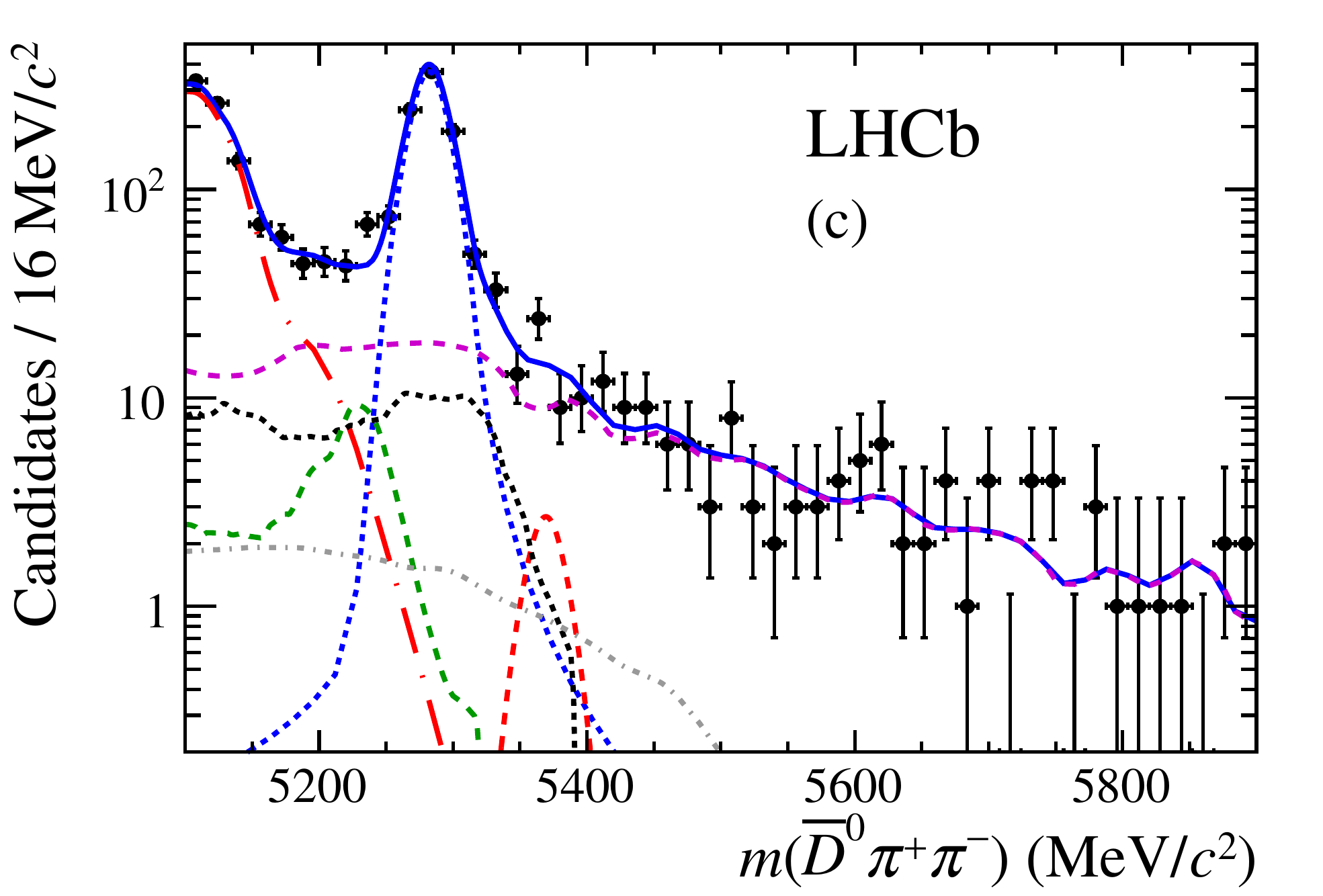}
\includegraphics[scale=0.39]{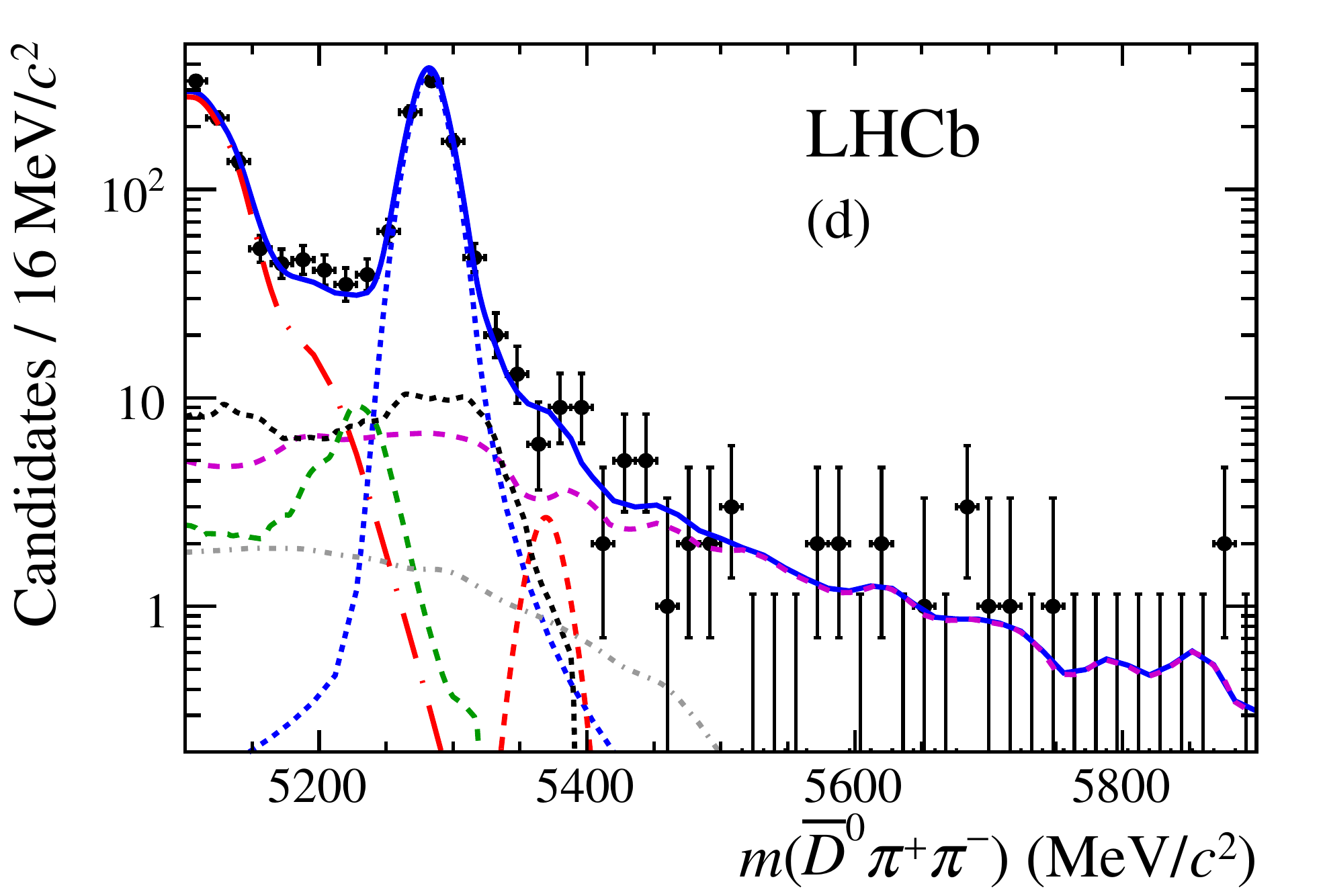}
\includegraphics[scale=0.39]{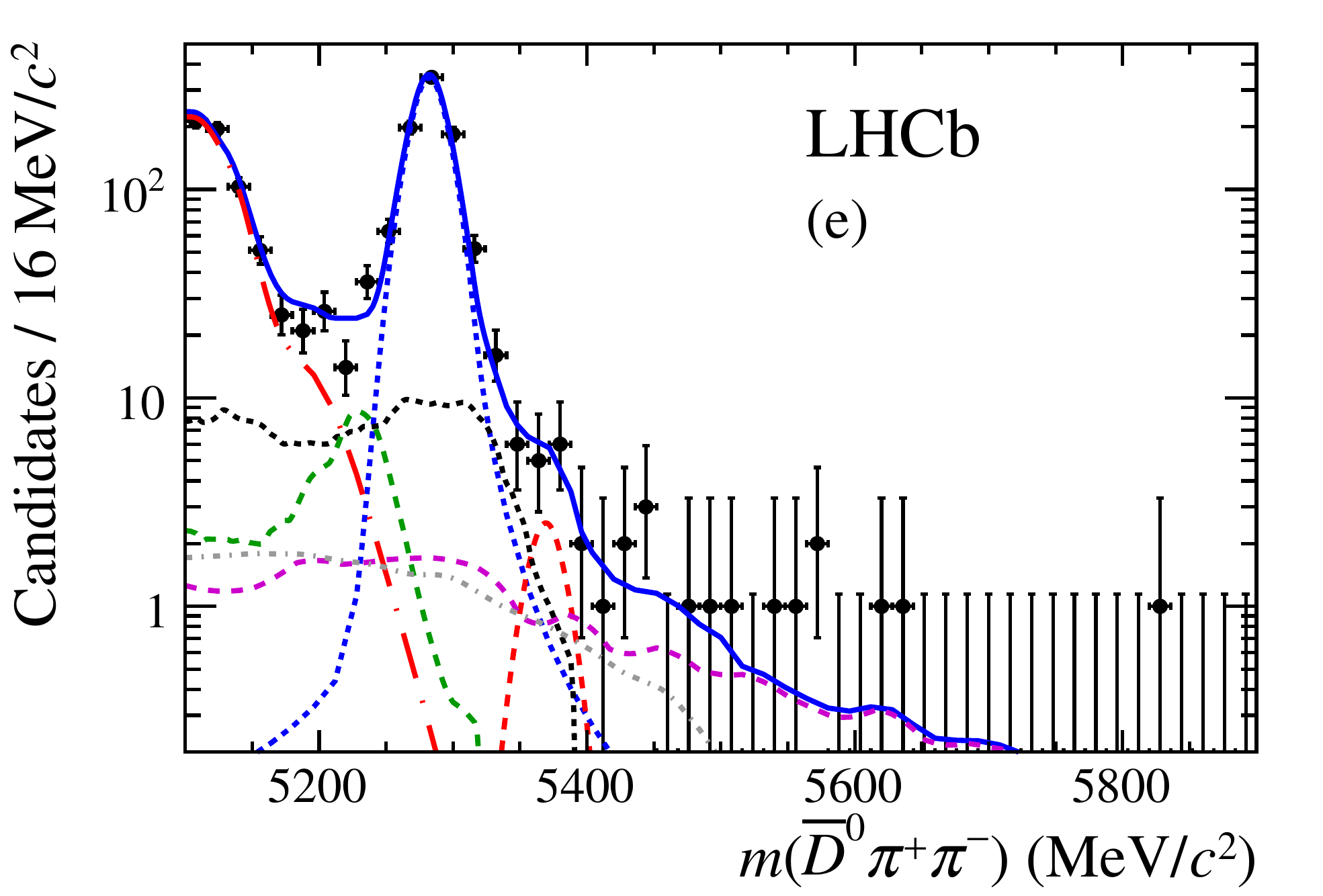}
\includegraphics[scale=0.39]{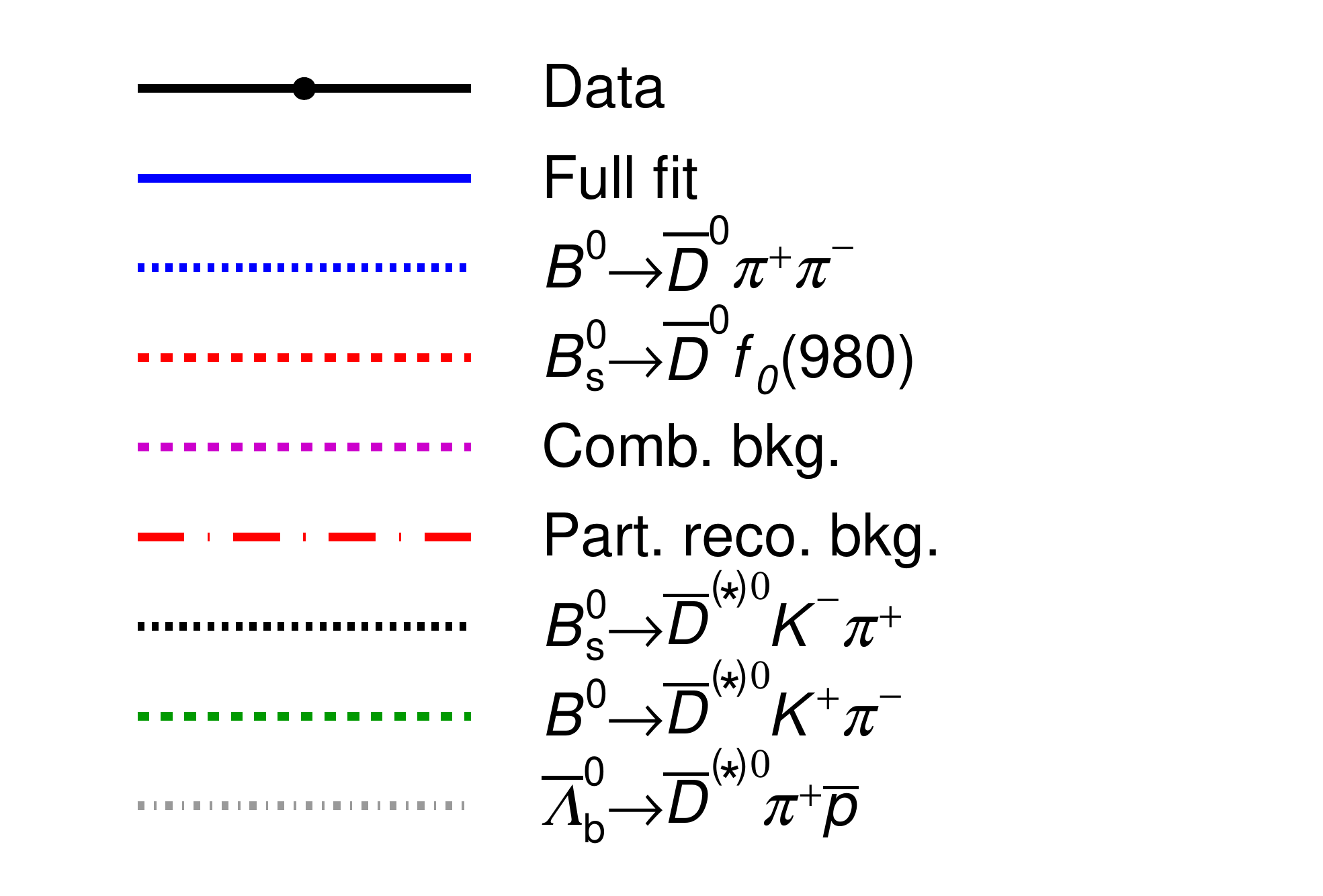}
\caption{\small
  Invariant mass distribution of candidates in the $\Dzb f_0(980)$ data sample with fit results overlaid, shown on a logarithmic scale.
  The components are as detailed in the legend. The labels (a) to (e) show the NN bins with increasing purity.
  The NN binning scheme is described in Sec.~\ref{sec:fitting}.
}
\label{fig:sigfit}
\end{figure}

The fit to the $\Dzb f_0(980)$ subsample includes the same components as the $\Bd\to\Dzb\pip\pim$ fit, with the addition of a second signal component to account for the possible presence of both $\Bd$ and $\Bs$ decays. 
The mass difference between the $\Bd$ and $\Bs$ mesons is fixed to the known value~\cite{PDG2014}.
The shapes for the $\Bd$ and $\Bs$ components are otherwise identical in both invariant mass and NN output.
The following parameters are fixed to the values found in the $\Bd\to\Dzb\pip\pim$ fit: the fractional yields $f_i$ for the signal and partially reconstructed background components; the relative normalisation of the two CB functions; the ratio of widths of the CB functions; the fraction of the partially reconstructed background from $\Bd\to\Dstarzb\pip\pim$ decays and the offset parameter of the partially reconstructed background.
In addition, the relative yields of the misidentified background components from $\Bd\to\DorDstarzb \Kp\pim$ and $\Bs\to\DorDstarzb \Km\pip$ decays are fixed to the expected value~\cite{LHCb-PAPER-2013-022}.
The remaining 14 parameters are: the yields for $\Bs\to\Dzb f_0(980)$ decays, $\Bd\to\Dzb\pip\pim$ decays, combinatorial and partially reconstructed backgrounds and for the $\Bds\to\DorDstarzb K\pi$ and $\Lb\to\DorDstarz p\pim$ misidentified backgrounds; the fractional yields of the combinatorial background in NN output bins, the exponential slope parameter of the combinatorial background and the fraction of the combinatorial background from $\Bu\to\Dstarzb\pip$ decays; and the signal peak position and core width.
Results of this fit are shown in Fig.~\ref{fig:sigfit}.

The yields from the fits to the $\Dzb f_0(980)$ and $\Dzb\pip\pim$ data samples are summarised in Table~\ref{tab:yields}. 
In total, $29 \pm 17$ $\Bs\to\Dzb f_0(980)$ decays are found, with a statistical significance of $2.2\,\sigma$ obtained from $\sqrt{-2 \; \Delta \ln {\cal L}}$, where $\Delta \ln {\cal L}$ is the change in log likelihood from the value obtained in a fit with zero signal yield.

\begin{table}[!tb]
\centering
\caption{\small
  Yields from the fit to the $\Dzb\pip\pim$ and $\Dzb f_0(980)$ samples.}
\label{tab:yields}
\begin{tabular}{lr@{$\;\pm\;$}lr@{$\;\pm\;$}l}
\hline \\ [-2.5ex]
 & \multicolumn{2}{c}{$\Dzb\pip\pim$} & \multicolumn{2}{c}{$\Dzb f_0(980)$} \\ [0.2ex]
\hline \\ [-2.5ex]
$\Bd\to\Dzb\pip\pim$ & $42\,636$ & $362$ & $3\,998$ & $87$ \\
$\Bs\to\Dzb f_0(980)$ & \multicolumn{2}{c}{---} & $\phantom{111}29$ & $17$ \\ [0.2ex]
Combinatorial & $90\,150$ & $481$ & $11\,064$ & $145$ \\
Partially reconstructed & $50\,950$ & $493$ & $3\,759$ & $88$ \\
$\Bds\to\DorDstarzb K\pi$ & $9\,225$ & $504$ & $852$ & $128$ \\
$\Lb \to\DorDstarz p\pim$ & $4\,923$ & $415$ & $154$ & $135$ \\
\hline
\end{tabular}
\end{table}

\section{Systematic uncertainties}
\label{sec:systematics}

The systematic uncertainties that affect the ratio of branching fractions are summarised in Table~\ref{tab:systs}.
Various effects contribute to the systematic uncertainties on the invariant mass fit and efficiencies, as described below.

The tail parameters of the signal components for $\Bs\to\Dzb f_0(980)$ and $\Bd\to\Dzb\pip\pim$ decays are varied within the uncertainties from the fit to simulated events. 
For the $\Bs\to\Dzb f_0(980)$ fit, the relative normalisation and ratio of widths of the CB functions are varied according to the uncertainties from the fit to the $\Bd\to\Dzb\pip\pim$ mode. 
Combined in quadrature these contribute $11.5\,\%$ to the systematic uncertainty.
The systematic uncertainty from assuming that the NN response is identical for $\Bs\to\Dzb f_0(980)$ and $\Bd\to\Dzb\pip\pim$ decays is evaluated by correcting the fractional yields found in the $\Bd\to\Dzb\pip\pim$ fit by the ratio of fractional yields found in simulated samples.
This contributes $0.3\,\%$ to the systematic uncertainty.

A second-order polynomial function is used to replace the exponential shape for the combinatorial background in both fits, giving a systematic uncertainty of $8.4\,\%$.
Varying the smoothing of the non-parametric shape for $\Bu\to\Dstarzb\pip$ decays gives the largest source of systematic uncertainty of $23.1\,\%$; the size of this effect is determined by that of the simulated background sample.
The fractional yields of the $\Bu\to\Dstarzb\pip$ component of the combinatorial background are fixed to the values found in simulation, rather than using the same fractional yields as the rest of the combinatorial background. 
This leads to a systematic uncertainty of $1.0\,\%$.

The smoothing of the non-parametric functions for $\Bd\to\Dstarzb\pip\pim$ and $\Bu\to\Dzb\pip\pim\pip$ is varied in both fits. 
Additionally, in the $\Bs\to\Dzb f_0(980)$ sample, the relative normalisation of the shapes is varied within uncertainties from the value found in the $\Bd\to\Dzb\pip\pim$ fit.
Combined in quadrature these contribute $6.2\,\%$ to the systematic uncertainty.
Allowing the fractional yields of the partially reconstructed background to vary in the $\Dzb f_0(980)$ fit, instead of being fixed to values found in the $\Dstarzb\pip\pim$ fit, contributes $3.1\,\%$ to the systematic uncertainty.

The misidentified background shapes are also varied by changing the smoothing applied to the non-parametric function. 
Additionally, the simulation is not reweighted to the known phase-space distributions and the relative normalisation of the $\Bds\to\DorDstarzb K\pi$ shapes is varied within uncertainties. 
Together these contribute $6.7\,\%$ to the systematic uncertainty.
Corrections, derived from simulation, are applied to the fractional yields for the misidentified backgrounds, which are assumed to behave like signal decays in the default fit. 
The sum in quadrature of the individual contributions gives a systematic uncertainty of $8.5\,\%$.

Potential biases in the fit procedure are investigated using an ensemble of pseudoexperiments.
Each of the pseudoexperiments is fitted with the same fit model used to describe the data samples.
This study shows that the fit is stable and well behaved and that the associated systematic uncertainty is negligible. 

The uncertainty on the ratio of reconstruction and selection efficiencies for the $\Bs\to\Dzb f_0(980)$ and $\Bd\to\Dzb\pip\pim$ final states contributes $2.5\,\%$ to the systematic uncertainty. 
This includes statistical uncertainty from the sizes of the simulated samples as well as effects related to the choice of binning in kinematic variables in the evaluation of the PID efficiency and potential differences in the response of the hardware trigger.
The simulated sample of $\Bs\to\Dzb f_0(980)$ decays is generated using a relativistic Breit--Wigner function with a width of $70\mev$ for the $f_0(980)$ meson. 
The true lineshape of the $f_0(980)$ meson can differ from the assumed shape in a process-dependent way, which can affect the fraction of $f_0(980)\to\pip\pim$ decays that fall inside the selected $m(\pip\pim)$ window. 
No systematic uncertainty is assigned due to this choice of $f_0(980)$ lineshape.
Other possible sources of uncertainty on the ratio of efficiencies are negligible.

The limited knowledge of the ratio of fragmentation fractions, $f_s/f_d = 0.259 \pm 0.015$~\cite{fsfd}, contributes $5.8\,\%$ to the systematic uncertainty.
Combining all of the above sources in quadrature, the total systematic uncertainty on the ratio of branching fractions is found to be $30.7\,\%$.

\begin{table}[!tb]
\centering
\caption{Summary of systematic uncertainties on the ratio of branching fractions.}
\label{tab:systs}
\begin{tabular}{lr}
\hline
Source & Value\phantom{.} \\
\hline
Signal shapes & $11.5\,\%$ \\
Combinatorial background shapes & $24.6\,\%$ \\
Partially reconstructed background shapes & $6.9\,\%$ \\
Misidentified background shapes & $10.8\,\%$ \\
Efficiencies & $2.5\,\%$ \\
Fragmentation fraction $f_s/f_d$ & $5.8\,\%$ \\
\hline
Total & $30.7\,\%$ \\
\hline
\end{tabular}
\end{table}

\section{Results and summary}
\label{sec:results}

The relative branching fraction of $\Bs\to\Dzb f_0(980)$ and $\Bd\to\Dzb\pip\pim$ decays is determined by correcting the ratio of yields for the relative efficiencies and fragmentation fractions, as shown in Eq.~(\ref{eqn:bf}).
The total efficiencies are found to be $\epsilon(\Bs\to\Dzb f_0(980)) = (0.76 \pm 0.02)\,\%$ and $\epsilon(\Bd\to\Dzb\pip\pim) = (0.57 \pm 0.02)\,\%$.
These values include contributions from the LHCb detector acceptance and from selection, trigger and PID requirements. 
The selection and trigger efficiencies are calculated from simulated samples with data-driven corrections applied.
The PID efficiency is measured using a control sample of $\Dstarm\to\Dzb\pim,\,\Dzb\to\Kp\pim$ decays.
Variation of the $\Bd\to\Dzb\pip\pim$ efficiency over the Dalitz plot is taken into account by weighting the simulation according to the observed Dalitz plot distribution~\cite{LHCb-PAPER-2014-070}.

Using Eq.~(\ref{eqn:bf}) the ratio of branching fractions is determined to be
$$
\frac{\mathcal{B}(\Bs\to\Dzb f_0(980))}{\mathcal{B}(\Bd\to\Dzb\pip\pim)} = (2.0 \pm 1.1 \pm 0.6) \times 10^{-3}\, ,
$$
where the first uncertainty is statistical and the second systematic. 
This result is obtained under the assumption that the $\Bs \to \Dzb \pip\pim$ decays proceed uniquely via the $f_0(980)$ resonance within the range $900 \mevcc < m(\pip\pim) < 1080 \mevcc$; no systematic uncertainty is assigned due to this assumption. 
The result can be converted into an absolute branching fraction by multiplying by $\mathcal{B}(\Bd\to\Dzb\pip\pim) = (8.46 \pm 0.51) \times 10^{-4}$~\cite{LHCb-PAPER-2014-070} to give
$$
\mathcal{B}(\Bs\to\Dzb f_0(980)) = (1.7 \pm 1.0 \pm 0.5 \pm 0.1) \times 10^{-6} \, ,
$$
where the third uncertainty is from $\mathcal{B}(\Bd\to\Dzb\pip\pim)$.
Since the signal yield is not significant, upper limits of 
$$
\mathcal{B}(\Bs\to\Dzb f_0(980)) <  3.1~(3.4)\times 10^{-6}
$$
are set at $90\,\%$ ($95\,\%$) confidence level. 
The statistical likelihood curve obtained from the fit is convolved with a Gaussian function of width equal to the systematic uncertainty. 
The limits obtained are the values within which $90\,\%$ ($95\,\%$) of the integral of the likelihood in the physical region of non-negative branching fraction are contained.

In summary, a search for the $\Bs\to\Dzb f_0(980)$ decay has been performed using $3.0\invfb$ of $pp$ collision data recorded by the LHCb detector in 2011 and 2012. 
No significant signal is observed, and a limit is set on the branching fraction that is below the predicted value~\cite{Kim:2013ria}.
The small yield suggests that much larger data samples will be necessary in order to determine the angle $\gamma$ of the CKM unitarity triangle with $\Bs\to\Dzb f_0(980)$ decays.
Table~\ref{tab:results} shows the current experimental status of measurements of the $\Bds \to \Dzb f_0(500)$ and $\Bds \to \Dzb f_0(980)$ branching fractions.
The pattern of branching fractions is very different to that for the $\Bds \to \jpsi f_0(500)$ and $\Bds \to \jpsi f_0(980)$ modes~\cite{LHCb-PAPER-2012-005,LHCb-PAPER-2012-045,LHCb-PAPER-2013-069,LHCb-PAPER-2014-012}.
These results may provide insight into the substructure of the scalar mesons.

\begin{table}[!tb]
\caption{\small
  Results for branching fractions for $\Bds \to \Dzb f_0(500)$ and $\Bds \to \Dzb f_0(980)$ decays.
  All quoted results correspond to branching fraction for $f_0 \to \pip\pim$ of $100\,\%$.
  There is no experimental result for ${\cal B}(\Bs \to \Dzb f_0(500))$.
}
\label{tab:results}
\centering
\begin{tabular}{ccc}
  \hline \\ [-2.5ex]
  & ${\cal B}(\Bz \to \Dzb f_0)$~\cite{LHCb-PAPER-2014-070} & ${\cal B}(\Bs \to \Dzb f_0)$ \\
  \hline
  $f_0(500)$ & $(11.2 \pm 0.8 \pm 0.5 \pm 2.1 \pm 0.5) \times 10^{-5}$ & --- \\
  $f_0(980)$ & $(1.34 \pm 0.25 \pm 0.10 \pm 0.46 \pm 0.06) \times 10^{-5}$ & $(1.7 \pm 1.0 \pm 0.5 \pm 0.1) \times 10^{-6}$ \\
  \hline
\end{tabular}
\end{table}

\section*{Acknowledgements}

\noindent We express our gratitude to our colleagues in the CERN
accelerator departments for the excellent performance of the LHC. We
thank the technical and administrative staff at the LHCb
institutes. We acknowledge support from CERN and from the national
agencies: CAPES, CNPq, FAPERJ and FINEP (Brazil); NSFC (China);
CNRS/IN2P3 (France); BMBF, DFG, HGF and MPG (Germany); INFN (Italy); 
FOM and NWO (The Netherlands); MNiSW and NCN (Poland); MEN/IFA (Romania); 
MinES and FANO (Russia); MinECo (Spain); SNSF and SER (Switzerland); 
NASU (Ukraine); STFC (United Kingdom); NSF (USA).
The Tier1 computing centres are supported by IN2P3 (France), KIT and BMBF 
(Germany), INFN (Italy), NWO and SURF (The Netherlands), PIC (Spain), GridPP 
(United Kingdom).
We are indebted to the communities behind the multiple open 
source software packages on which we depend. We are also thankful for the 
computing resources and the access to software R\&D tools provided by Yandex LLC (Russia).
Individual groups or members have received support from 
EPLANET, Marie Sk\l{}odowska-Curie Actions and ERC (European Union), 
Conseil g\'{e}n\'{e}ral de Haute-Savoie, Labex ENIGMASS and OCEVU, 
R\'{e}gion Auvergne (France), RFBR (Russia), XuntaGal and GENCAT (Spain), Royal Society and Royal
Commission for the Exhibition of 1851 (United Kingdom).

\ifx\mcitethebibliography\mciteundefinedmacro
\PackageError{LHCb.bst}{mciteplus.sty has not been loaded}
{This bibstyle requires the use of the mciteplus package.}\fi
\providecommand{\href}[2]{#2}

\clearpage

\clearpage
%%%%%%%%%%%%%%%%%%%%%%%%%%%%%%%%%%%%%%%%%%
\centerline{\large\bf LHCb collaboration}
\begin{flushleft}
\small
R.~Aaij$^{38}$, 
B.~Adeva$^{37}$, 
M.~Adinolfi$^{46}$, 
A.~Affolder$^{52}$, 
Z.~Ajaltouni$^{5}$, 
S.~Akar$^{6}$, 
J.~Albrecht$^{9}$, 
F.~Alessio$^{38}$, 
M.~Alexander$^{51}$, 
S.~Ali$^{41}$
% ...

{\footnotesize \it
% ...
$ ^{5}$Clermont Universit\'{e}, Universit\'{e} Blaise Pascal, CNRS/IN2P3, LPC, Clermont-Ferrand, France\\
$ ^{6}$CPPM, Aix-Marseille Universit\'{e}, CNRS/IN2P3, Marseille, France\\
% ...
$ ^{9}$Fakult\"{a}t Physik, Technische Universit\"{a}t Dortmund, Dortmund, Germany\\
% ...
$ ^{37}$Universidad de Santiago de Compostela, Santiago de Compostela, Spain\\
$ ^{38}$European Organization for Nuclear Research (CERN), Geneva, Switzerland\\
% ...
$ ^{41}$Nikhef National Institute for Subatomic Physics, Amsterdam, The Netherlands\\
% ...
$ ^{46}$H.H. Wills Physics Laboratory, University of Bristol, Bristol, United Kingdom\\
% ...
$ ^{51}$School of Physics and Astronomy, University of Glasgow, Glasgow, United Kingdom\\
$ ^{52}$Oliver Lodge Laboratory, University of Liverpool, Liverpool, United Kingdom\\
% ...
}
\end{flushleft}
%%%%%%%%%%%%%%%%%%%%%%%%%%%%%%%%%%%%%%%%%%

\end{document}